\documentclass[aps,prd,groupedaddress,showpacs,nofootinbib]{revtex4}
\usepackage{graphicx,amssymb,amsmath}
\usepackage[english]{babel}

\newcommand{\be}{\begin{equation}}\newcommand{\ee}{\end{equation}}
\newcommand{\bea}{\begin{eqnarray}}\newcommand{\eea}{\end{eqnarray}}
\newcommand{\beaa}{\begin{eqnarray}}\newcommand{\eeaa}{\end{eqnarray}}
\newcommand{\ba}{\begin{array}}\newcommand{\ea}{\end{array}}
\newcommand{\bit}{\begin{itemize}}\newcommand{\eit}{\end{itemize}}
\newcommand{\ben}{\begin{enumerate}}\newcommand{\een}{\end{enumerate}}

\def\lan{\langle}
\def\lf{\left}

\def\ran{\rangle}

\def\ri{\right}

\def\al{\alpha}\def\bt{\beta}

\def\te{\theta}

\def\1{{_{1}}}\def\2{{_{2}}}

\begin{document}

\title{Cosmological effects of neutrino mixing}

\author{M. Blasone${}^{\flat}$, A.Capolupo${}^{\natural}$, S.Capozziello${}^{\sharp}$, G.Vitiello${}^{\flat}$}


\affiliation{${}^{\flat}$ Dipartimento di Matematica e
Informatica,
 Universit\`a di Salerno and Istituto Nazionale di Fisica Nucleare,
 Gruppo Collegato di Salerno, 84100 Salerno, Italy,
\\
${}^{\natural}$ Department of Physics and Astronomy,
University of Leeds, Leeds LS2 9JT UK,
\\   ${}^{\sharp}$ Dipartimento di Scienze Fisiche, Universit\`a di Napoli "Federico II" and INFN Sez. di Napoli,
Compl. Univ. Monte S. Angelo, Ed.N, Via Cinthia, I-80126 Napoli,
Italy.}


\vspace{2mm}

\begin{abstract}

We report on the recent result that a contribution to the dark
energy can be achieved by the vacuum condensate induced by
neutrino mixing phenomenon.

\end{abstract}


\maketitle


The energy content of the neutrino-antineutrino pair condensate in
the vacuum can have effects on cosmic scale \cite{Blasone:2004yh}
and, in particular, can be interpreted as dynamically evolving
dark energy \cite{Capolupo:2006et} that, at present epoch, assumes
the behavior and the value of the observed cosmological constant.
This result is obtained by studying the neutrino mixing phenomenon
\cite{Pontecorvo:1957cp} - \cite{Giunti:1991ca} in the framework
of the quantum field theory (QFT) formalism, see \cite{BV95} -
\cite{Blasone:2006jx} and refs. therein cited. Indeed, the unitary
inequivalence between the massive neutrino vacuum and the flavor
vacuum leads to new oscillation formulas and to a non-zero
contribution to the vacuum energy.



We summarize the main features of the QFT formalism for the
neutrino mixing (for a detailed review see
\cite{Capolupo:2004av}). The Pontecorvo mixing transformations for
two Dirac neutrino fields are
\begin{eqnarray} \label{mix1}
\nu_{e}(x) &=&\nu_{1}(x)\,\cos\theta + \nu_{2}(x)\,\sin\theta
\\ [2mm]
\label{mix2} \nu_{\mu}(x) &=&-\nu_{1}(x)\,\sin\theta +
\nu_{2}(x)\,\cos\theta \;,\end{eqnarray}
where $\nu_{e}(x)$ and $\nu_{\mu}(x)$ are the fields with definite
flavors, $\theta$ is the mixing angle and $\nu_1$ and $\nu_2$ are
the fields with definite masses $m_{1} \neq m_{2}$:
\bea\label{freefi}
 \nu _{i}(x)=\frac{1}{\sqrt{V}}{\sum_{{\bf k} ,
r}} \left[ u^{r}_{{\bf k},i}\, \al^{r}_{{\bf k},i}(t) +
v^{r}_{-{\bf k},i}\, \bt^{r\dag}_{-{\bf k},i}(t) \ri] e^{i {\bf
k}\cdot{\bf x}},\qquad \, \qquad i=1,2, \eea with  $ \al_{{\bf
k},i}^{r}(t)=\al_{{\bf k},i}^{r}\, e^{-i\omega _{k,i}t}$, $
\bt_{{\bf k},i}^{r\dag}(t) = \bt_{{\bf k},i}^{r\dag}\,
e^{i\omega_{k,i}t},$ and $ \omega _{k,i}=\sqrt{{\bf k}^{2} +
m_{i}^{2}}.$ The operators $\alpha ^{r}_{{\bf k},i}$ and $ \beta
^{r }_{{\bf k},i}$, $ i=1,2 \;, \;r=1,2$ annihilate the vacuum
state $|0\rangle_{1,2}\equiv|0\rangle_{1}\otimes |0\rangle_{2}$.

The mixing transformation Eqs.(\ref{mix1}), (\ref{mix2}) can be
written as \cite{BV95}:
\bea \label{mixG} \nu_{e}^{\alpha}(x) = G^{-1}_{\bf \te}(t)\;
\nu_{1}^{\alpha}(x)\; G_{\bf \te}(t)
\\
 \nu_{\mu}^{\alpha}(x)
= G^{-1}_{\bf \te}(t)\; \nu_{2}^{\alpha}(x)\; G_{\bf \te}(t) \eea
where $G_{\bf \te}(t)$ is the mixing generator that, at finite
volume, is an unitary operator, $G^{-1}_{\bf \te}(t)=G_{\bf
-\te}(t)=G^{\dag}_{\bf \te}(t)$, preserving the canonical
anticommutation relations. $G^{-1}_{\bf \te}(t)$ maps the Hilbert
spaces for $\nu_1$ and $\nu_2$ fields ${\cal H}_{1,2}$ to the
Hilbert spaces for the flavor fields ${\cal H}_{e,\mu}$. In
particular, we have $ |0(t) \rangle_{e,\mu} = G^{-1}_{\bf
\te}(t)\; |0 \rangle_{1,2}\; $, where $|0 \rangle_{e,\mu}$ is the
vacuum for ${\cal H}_{e,\mu}$, which is referred to as the flavor
vacuum. In the infinite volume limit the flavor vacuum $|0(t)
\rangle_{e,\mu}$ turns out to be unitarily inequivalent to the
vacuum for the massive neutrinos $|0 \rangle_{1,2}$ \cite{BV95}.
For a discussion on QFT favor states see \cite{BV95} -
\cite{Blasone:2006jx} and \cite{paschos1}.

The flavor annihilators, relative to the fields $\nu_{e}(x)$ and
$\nu_{\mu}(x)$, are
\begin{eqnarray}\label{flavannich}
\alpha _{{\bf k},\sigma}^{r}(t) &\equiv &G^{-1}_{\bf
\te}(t)\;\alpha _{{\bf k},i}^{r}(t)\;G_{\bf \te}(t),
\\
\beta _{{\bf k},\sigma}^{r}(t) &\equiv &G^{-1}_{\bf \te}(t)\;\beta
_{{\bf k},i}^{r}(t)\;G_{\bf \te}(t)\,,
\end{eqnarray}
with  $(\sigma,i)=(e,1), (\mu,2)$. In the reference frame such
that ${\bf k}=(0,0,|{\bf k}|)$ they are explicitly:
 \bea\label{annihilator}
\alpha^{r}_{{\bf k},e}(t)&=&\cos\theta\;\alpha^{r}_{{\bf
k},1}(t)\;+\;\sin\theta\;\left( |U_{{\bf k}}|\; \alpha^{r}_{{\bf
k},2}(t)\;+\;\epsilon^{r}\; |V_{{\bf k}}|\; \beta^{r\dag}_{-{\bf
k},2}(t)\right) \eea and similar for $\alpha^{r}_{{\bf k},\mu} $,
$\beta^{r}_{{\bf k},e}$, $ \beta^{r}_{{\bf k},\mu}$, where $
|U_{{\bf k}}| \equiv  u^{r\dag}_{{\bf k},i} u^{r}_{{\bf k},j} =
v^{r\dag}_{-{\bf k},i} v^{r}_{-{\bf k},j}\,, \quad  |V_{{\bf k}}|
\equiv  \epsilon^{r}\; u^{r\dag}_{{\bf k},1} v^{r}_{-{\bf k},2} =
-\epsilon^{r}\; u^{r\dag}_{{\bf k},2} v^{r}_{-{\bf k},1} $ with
$i,j = 1,2$ and $ i \neq j$. We have $ |U_{{\bf k}}|^{2}+|V_{{\bf
k}}|^{2}=1$. Extension of the QFT formalism to three neutrinos has
been also done \cite{BV95} - \cite{Blasone:2006jx}. In the present
report we consider only the two neutrino case.

The number of condensed neutrinos is given by
\bea \label{density} _{e,\mu}\langle 0| \al_{{\bf k},i}^{r \dag}
\al^r_{{\bf k},i} |0\rangle_{e,\mu}\,= \;_{e,\mu}\langle 0|
\bt_{{\bf k},i}^{r \dag} \bt^r_{{\bf k},i}
|0\rangle_{e,\mu}\;=\;\sin^{2}\te\; |V_{{\bf k}}|^{2} \;, \quad
i=1,2.
 \eea
The Bogoliubov coefficient $|V_{{\bf k}}|^{2}$ is zero for $m_1 =
m_2$, it has a maximum at $|{\bf k}|^2 = m_{1} m_{2}$ and goes to
zero for large momenta (i.e. for $|{\bf k}|^2\gg m_{1} m_{2}$ ) as
$|V_{{\bf k}}|^2 \approx \frac{(\Delta m)^2}{4 k^2}$. It acts as a
``form factor'' (in the $\bf k$ space) controlling the neutrino
vacuum condensate. The mixing of neutrinos contributes to the dark
energy exactly because of the non-zero value of $|V_{\bf k}|^2$
\cite{Blasone:2004yh}: its behavior at very high momenta, together
with the Lorentz invariance of the vacuum condensate at the
present time, leads to a contribution of the dark energy that has
the very tiny value of the cosmological constant
\cite{Capolupo:2006et}.

Let us calculate the contribution $\rho_{vac}^{mix}$ of the
neutrino mixing to the vacuum energy density. The calculation is
performed in a Minkowski space-time but it can be easily extended
to curved space-times. The treatment here presented is a good
approximation in the present epoch of that in FRW space-time
since, at present epoch, the characteristic oscillation length of
the neutrino is much smaller than the universe curvature radius.
When computations are carried on in a curved background, the
mixing mechanism gives a time dependent dark energy, leading,
however, to the same final result we obtain in the flat
space-time.

The Lorentz invariance of the vacuum implies that the vacuum
energy-momentum tensor density is equal to zero: ${\cal
T}_{\mu\nu}^{vac} = \lan 0 |:{\cal T}_{\mu\nu}:| 0\ran= 0$,
($:...:$ denotes the normal ordering). The (0,0) component of
${\cal T}_{\mu\nu}(x) $ for the fields $\nu_1$ and $\nu_2$ is $
:{\cal T}_{00}(x): = \frac{i}{2}:\left({\bar
\Psi}_{m}(x)\gamma_{0} \stackrel{\leftrightarrow}{\partial}_{0}
\Psi_{m}(x)\right): $, where $\Psi_{m} = (\nu_1, \nu_2)^{T}$. In
terms of the annihilation and creation operators of fields
$\nu_{1}$ and $\nu_{2}$ the (0,0) component of the energy-momentum
tensor $ T_{00}=\int d^{3}x {\cal T}_{00}(x)$ is given by $\,
:T^{00}_{(i)}:= \sum_{r}\int d^{3}{\bf k}\,
\omega_{k,i}\lf(\al_{{\bf k},i}^{r\dag} \al_{{\bf k},i}^{r}+
\beta_{{\bf -k},i}^{r\dag}\beta_{{\bf -k},i}^{r}\ri),$ with
$i=1,2$. Note that $T^{00}_{(i)}$ is time independent.

In the early universe epochs, when the Lorentz invariance of the
vacuum condensate is broken,
 $\rho_{vac}^{mix}$ presents also space-time dependent condensate
 contributions \cite{Capolupo:2006et}.
Then $\rho_{vac}^{mix}$ and the contribution of the neutrino
mixing to the vacuum pressure
 $ p_{vac}^{mix}$ are given respectively by:
 \bea
\rho_{vac}^{mix}= -\frac{1}{V}\; \eta_{00} \; {}_{e,\mu}\lan 0
|\sum_{i} :T^{jj}_{(i)}:| 0\ran_{e,\mu}\,,
\\
p_{vac}^{mix}= -\frac{1}{V}\; \eta_{jj} \; {}_{e,\mu}\lan 0
|\sum_{i} :T^{jj}_{(i)}:| 0\ran_{e,\mu}\,,
 \eea
where $ :T^{jj}_{(i)}:= \sum_{r}\int d^{3}{\bf k}\, \frac{k^j
k^j}{\;\omega_{k,i}}\lf(\al_{{\bf k},i}^{r\dag} \al_{{\bf
k},i}^{r}+ \beta_{{\bf -k},i}^{r\dag}\beta_{{\bf -k},i}^{r}\ri)$
(no summation on the index $j$ is intended). This implies that the
adiabatic index is $w = p_{vac}^{mix}/ \rho_{vac}^{mix} \simeq
1/3$ when the cut-off is chosen to be $K \gg m_{1}, m_{2} $
\cite{Capolupo:2006et}.

At the present epoch, the breaking of the Lorentz invariance is
negligible and then
 $\rho_{vac}^{mix}$ comes from space-time independent condensate contributions
 (i.e. the contributions carrying a non-vanishing $\partial_{\mu} \sim k_{\mu}=(\omega_{k},k_{j}) $
are missing).
 Thus, the energy-momentum density tensor of the vacuum condensate
is given by \bea {}_{e,\mu}\lan 0 |:T_{\mu\nu}:| 0\ran_{e,\mu} =
\eta_{\mu\nu}\;\sum_{i}m_{i}\int
\frac{d^{3}x}{(2\pi)^3}\;{}_{e,\mu}\lan 0 |:\bar{\nu
}_{i}(x)\nu_{i}(x):| 0\ran_{e,\mu}\, =
\eta_{\mu\nu}\;\rho_{\Lambda}^{mix}.
 \eea

Since $\eta_{\mu\nu} = diag (1,-1,-1,-1)$ and, in a homogeneous
and isotropic universe, the energy-momentum tensor is $T_{\mu\nu}
= diag (\rho\,,p\,,p\,,p\,)$, then the state equation is
$\rho_{\Lambda}^{mix} = -p_{\Lambda}^{mix}$. This means that
$\rho_{\Lambda}^{mix}$ contributes today to the dynamics of the
universe by a
 cosmological constant behavior {\cite{Capolupo:2006et}. We
obtain
\bea\label{cost} \rho_{\Lambda}^{mix}= \frac{2}{\pi}
\sin^{2}\theta \int_{0}^{K} dk \,
k^{2}\lf[\frac{m_{1}^{2}}{\omega_{k,1}}+\frac{m_{2}^{2}}
{\omega_{k,2}}\ri] |V_{\bf k}|^{2}, \eea
where $K$ is the cut-off on the momenta. Notice the presence of
the $|V_{\bf k}|^{2}$ factor in the integrand (\ref{cost}):
$\rho_{\Lambda}^{mix}$ would be zero for $|V_{\bf k}|^{2}= 0$ for
any $|\bf k|$, as it is in the usual quantum mechanical Pontecorvo
formalism.

For neutrino masses of order of $10^{-3}eV$ and $\Delta m^{2}
\approx 10^{-5} eV^{2}$ we have $\rho_{\Lambda}^{mix} = 2.9 \times
10^{-47}GeV^{4}$, which is the observed dark energy value, even
for a value of the cut-off of order of the Planck scale $K=10^{19}
GeV$.

In conclusion, the vacuum condensate generated by neutrino mixing
can be interpreted as an evolving dark energy that, at present
epoch, behaves as the cosmological constant, giving rise to its
observed value. The result is achieved even when a cut-off $K$ of
the order of Planck scale is considered. Such a result links
together dark energy with the neutrino masses.

So far we have considered the case of mixing of lighter neutrino
generations ($\nu_1$ and $\nu_2$). When the three generation
mixing is considered we obtain a value for $\rho_{\Lambda}^{mix}$
which differs of about $5$ order of magnitude from the observed
dark energy value. We observe that even in such a case our result
represents a dramatic improvement with respect to the $123$ order
of magnitude of disagreement obtained in other theoretical
approaches, which makes our treatment worth to be studied and
discussed. Moreover, our approach to the dark energy problem does
not need the {\it ad hoc} introduction of  exotic fields and
particles. Finally, we mention that one plausibility argument
which is put forward in the literature \cite{mavromatos} to
justify the exclusion of heavier neutrinos from the dark energy
contribution is based on the abundance of electrons in the galaxy,
implying the dominance of mixing and oscillations of lighter
neutrinos. However, a deeper theoretical understanding of such an
exclusion is needed and we are presently working on this.

We thank Arttu Rajantie and the organizers of PASCOS 07 for giving
us the opportunity to present this work at the PASCOS Conference.
We also acknowledge INFN for financial support.





\end{document}